\documentclass[12pt]{raa}
\usepackage{amsmath}            
\usepackage{graphicx,times}
\usepackage{natbib}

\newcommand{\aaa}{A\&A}
\newcommand{\mn}{MNRAS}

\definecolor{gray}{rgb}{0.75,0.75,0.75}
\definecolor{mygreen}{rgb}{0.2,0.7,0.2}
\definecolor{myyellow}{rgb}{0.7,0.5,0.05}

\begin{document}

\title{The color gradients of spiral disks in the Sloan Digital Sky Survey}

 \volnopage{ {\bf 2009} Vol.\ {\bf X} No. {\bf XX}, 000--000}
   \setcounter{page}{1}

\author{Cheng-Ze Liu\inst{1,2,3,4}
       \and Shi-Yin Shen\inst{1,3}, Zheng-Yi Shao\inst{1,3}, Rui-Xiang Chang\inst{1,3},
        Jin-Liang Hou\inst{1}
       \and Jun Yin\inst{1,2}
       \and Da-Wei Yang\inst{4}}



\institute{Key Laboratory for Research in Galaxies and
   Cosmology, Shanghai Astronomical Observatory, Chinese Academy of
   Sciences, 80 Nandan RD, Shanghai, 200030, China; czliu@shao.ac.cn\\
\and
   Graduate school of Chinese Academy of Sciences, 19A
    Yuquan Road, Beijing, 100049, China
\and
   Key Laboratory for Astrophysics, Shanghai, 200234, China
\and
   Department of Physics, Hebei Normal University, 113
    East Yuhua Road, Shijiazhuang, 050016, China.}



\abstract{We investigate the radial color gradients of galactic
disks using a sample of $\sim20,000$ face-on spiral galaxies
selected from the fourth data release of the Sloan Digital Sky
Survey (SDSS-DR4). We combine galaxies with similar concentration,
size and luminosity to construct $composite$ galaxies, and then
measure their color profiles by stacking the azimuthally averaged
radial color profiles of all the member galaxies. Except for the
smallest galaxies ($R_{50}<3~\rm kpc$), almost all galaxies show
negative disk color gradients with mean $g-r$ gradient
$\bar{G}_{gr}=-0.006$\,mag\,kpc$^{-1}$ and $r-z$ gradient
$\bar{G}_{rz}=-0.018$\,mag\,kpc$^{-1}$. The disk color gradients are
independent of the morphological types of galaxies and strongly
dependent on the disk surface brightness $\mu_{d}$, with lower
surface brightness galactic disks having steeper color gradients. We
quantify the intrinsic correlation between color gradients and
surface brightness as $G_{gr}=-0.011\mu_{d}+0.233$ and
$G_{rz}=-0.015\mu_{d}+0.324$. These quantified correlations provide
tight observational constraints on the formation and evolution
models of spiral galaxies. }

\authorrunning{C.~Z. Liu et al. }          
\titlerunning{Color gradients of face-on spiral galaxies}  
\maketitle


\section{INTRODUCTION}\label{sec:intro}

Radial color gradient of a spiral galaxy freezes its fossil
information of the star formation history along the disk and
therefore provides a strong constraint on the models of disk
formation and evolution. Earlier studies revealed that most of the
spiral galaxies show negative color gradients, i.e. the color is
gradually bluer outwards (e.g., \citealt{kim90, deJong96b, bell00,
gadotti01, moth02, macArthur04, taylor05}).

\citet{kim90} studied the radial color distributions of a sample of
103 face-on galaxies with different morphological types. They
concluded that about 40 percent of elliptical and lenticular
galaxies and about 75 percent of intermediate type spiral galaxies
show clear negative color gradients, while the latest type spiral
and irregular galaxies have positive gradients or no gradient at
all. For spiral galaxies, they found a weak correlation between the
color gradient and absolute magnitude, with more luminous spirals
having steeper gradients. However, in their work, the color gradient
was measured along the whole image of each galaxy, where the bulge
and disk components were not decomposed.

It is well established that a spiral galaxy can be decomposed into a
concentrated bulge plus an exponential disk \citep{dev59, simien90,
capaccioli92}. Therefore, the negative color gradient of a spiral
galaxy might be partially originated from the configuration of these
two components. To remove this effect and measure the color
gradients of these two components separately, a bulge-disk
decomposition is required. \citet{deJong96a, deJong96b, deJong96c}
made the bulge-disk decomposition for a sample of 86 face-on spiral
galaxies and investigated radial color distributions of disk
components in optical and near-infrared bands. He found that the
color of disk component also becomes bluer outwards, but no
significant correlations were found between the color gradients and
other structural parameters of galaxies, such as the color,
inclination, morphological type, central surface brightness,
scale-length, integrated magnitude and so on. However, the
decomposition is dependent on the models used. Different models used
could give different results. Alternatively, \citet{taylor05}
investigated the color gradients of the bulge dominated inner parts
and the disk dominated outer parts of 142 galaxies by simply
separating each galaxy at the effective radius $R_e$. For the disk
parts, they found that the early and intermediate type spiral
galaxies show weak negative color gradients, while the late type
spiral and irregular galaxies have positive color gradients.

Although the color gradient has been investigated extensively, its
quantitative definitions are still diverse in different studies.
There are mainly two branches of such definitions. One is the color
variation $\Delta CI$ as a function of scaled radius (e.g.,
scale-length $R_{d}$, half-light radius $R_{50}$), either in linear
space \citep{segalovitz75, vader88, kim90, taylor05} or in
logarithmic space \citep{tamura03, wu05, li05}. This kind of
definition implies that galaxies with different sizes have
similarities in the structure of color distribution. The other
branch is the $\Delta CI$ measured along the physical radius, e.g.
the color variation per kpc (\citealt{prugniel98, gadotti01,
moth02}), which could also be presented in either linear or
logarithmic space. Additionally, if one assumes that the galactic
disk in different wavebands could all be well modeled by exponential
profiles, the change of scale-length as a function of wavebands
could also be used to parameterize the color gradient
\citep{peletier94, deJong96b, cunow04}. Despite the diversity of the
definitions, the color gradient is a essential parameter that
quantify the systematical change of color profile of galaxy.

To derive the color profile of a galaxy, detailed photometric
analyses in different wavebands are required. Moreover, to get
robust conclusions on the correlations between color gradients and
other structural parameters, a large and homogenous sample of
galaxies is needed. Up to now, the number of galaxies with detailed
measurement of color profile is quite limited \citep{gadotti01,
pohlen06, munoz07, azzollini08}. On the other hand, the Sloan
Digital Sky Survey (SDSS, \citealt{york00}) has surveyed about a
quarter of the whole sky in 5 broad wavebands $ (u,g,r,i,z)$ and
provided a large catalogue of photometric data of galaxies
\citep{adelman07}. In the released catalogue, the surface brightness
profile of each galaxy has been measured in a series of fixed
concentric annuli, known as $profMean$\footnote{Before calculating
the surface brightness profile $profMean$, the foreground and
background objects have been deblended by the \texttt{FRAMES}
pipeline, see \citet{lupton01} and \cite{stoughton02}.} (see
\citealt{stoughton02} for detail, hereafter EDR). Due to the short
exposure time (about 54 second) of SDSS image,  the signal-to-noise
ratio (S/N) of $profMean$ data is limited and it is suggested to be
only used in the cumulative profile for individual galaxies in EDR
paper\footnote{ \cite{blanton03b} has used $profMean$ and
successfully fitted the surface brightness profile with a Sersic
model for each individual galaxy.}. However, the S/N could be
greatly improved by stacking the profiles ($profMean$) of a large
number of similar galaxies. The large amount of galaxies in SDSS
provides an opportunity to make subsamples with enough member
galaxies for such stacking. Comparing to the direct image stacking
method with contaminated sources carefully masked (e.g.
\citealt{zibetti04, deJong08, hathi08}), this kind of catalog data
stacking is more vulnerable to background contaminations, but is
much simpler and more suitable for the analysis of large data sets
\citep{azzollini08, bakos08}.

In this study, we use a large sample of face-on spiral galaxies
($\sim$ 20,000) selected from SDSS-DR4 \citep{adelman06} to
investigate their disk color gradients statistically. We divide
these spiral galaxies into subsamples with similar concentrations,
sizes and luminosities, and then construct $composite$ galaxies by
combining the color profiles of the member galaxies in each
subsample. After measuring the color gradients of disk components of
these $composite$ galaxies, we further explore possible correlations
between color gradients and structural parameters. The main aim of
this work is to find out which parameter is primarily  correlated
with the color gradient and discuss its physical implications.

The structure of this paper is as follows. In
Section~\ref{sec:data}, we describe our sample selection and the
building of $composite$ galaxies. The color gradients and their
correlations with other structural parameters are presented in
Section~\ref{sec:result}. Finally, we give a brief summary and make
discussions in Section~\ref{sec:dis}.

\section{DATA}\label{sec:data}

\subsection{Sampling}\label{sec:data-sample}

The galaxies used in this paper are drawn from the main galaxy
sample of the SDSS-DR4, which is a spectroscopic sample complete to
$r$-band Petrosian magnitudes $r_\textrm{P}<17.77$. We use galaxies
with $r_\textrm{P}<17.5$ and redshift $z>0.01$. The magnitude limit
excludes the faint objects, ensuring the structure of the galaxies
being nicely resolved by SDSS images \citep{desroches07}. The low
redshift limit eliminates the effect of the peculiar velocity on the
distance estimation.

\begin{figure}[h!!!]
 \centering
 \includegraphics[width=0.8\textwidth]{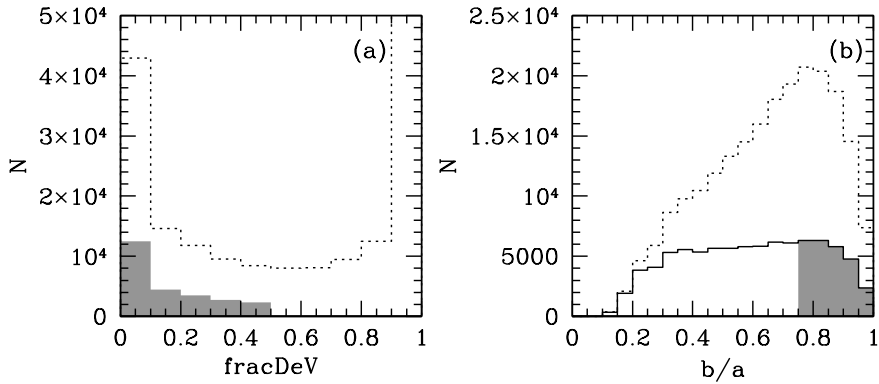}\\
\caption{The histograms of $fracDeV$ [left panel (a)] and axis ratio
$b/a$ [right panel (b)]. The dotted histograms show the
distributions of the initial sample with $r_\textrm{P}<17.5$ and
$z>0.01$ while the shadowed regions represent the distributions of
final sample of 25,517 face-on spiral galaxies ($r_\textrm{P}<17.5$,
$z>0.01$, $fracDeV<0.5$ and $b/a>0.75$). The solid histogram in
panel (b) shows the $b/a$ distribution of selected spiral galaxies
 ($fracDeV<0.5$), see text for detail.} \label{fig:hist_para}
\end{figure}

To select spiral galaxies, we adopt the criterion $fracDeV<0.5$ (in
$r$-band\footnote{In this paper, all the photometric parameters are
in $r$-band as default unless we specify the other band with a
subscript.}). The parameter $fracDeV$ indicates the likelihood of
the surface brightness profile that can be best modeled by a de
Vaucouleurs profile \citep{abazajian04} and has been widely used as
a galaxy morphology indicator (e.g., \citealt{bernardi05, chang06,
shao07, padilla08, zhong08}). \citet{strateva01} have shown that the
criterion $fracDeV<0.5$ can separate late type galaxies from early
type ones with a reliability $\sim90$ percent. We show the histogram
of $fracDeV$ of sample galaxies in the left panel of Fig. 1. As we
can see, for the galaxies with $fracDeV<0.5$, most of them peak at
$fracDeV\sim0$.

Aiming at the study of color gradients, we further restrict our
spirals to be only face-on galaxies. This restriction is based on
two considerations. One is that the internal dust attenuation in the
face-on case is minimized. The other is that the $profMean$, which
is measured in circular apertures, is more reliable for face-on
galaxies. We adopt the criterion $b/a>0.75$ to make such face-on
selection, where $b/a$ is the axis ratio derived from the
exponential profile fitting and given as \texttt{ab\_exp} in SDSS
catalog\footnote{Both $fracDeV$ and \texttt{ab\_exp} are point
spread function corrected fitting parameters in the SDSS catalog.}
(see EDR paper). We show the histograms of $b/a$ of our sample
galaxies in the right panel of Fig.1. The dotted and solid
histograms represent the distributions of galaxies before and after
the morphological selection ($fracDeV<0.5$), respectively. As
expected,  the $b/a$ shows a flat distribution and parameterizes the
inclination of spiral disk very well \citep{lambas92, shao07}.
According to the study of \citet{shao07}, the axis ratio
$b/a\sim0.75$ corresponds to inclination angle $i\sim41^\circ$ (see
fig.~3 of their paper).

Finally, we select a magnitude limit sample of 25,517 face-on spiral
galaxies, whose distributions of $fracDeV$ and $b/a$ are shown as
shadow regions in Fig.~\ref{fig:hist_para}.

\subsection{Binning and composing galaxies }\label{sec:data-cg-bin}

In this section, we describe how to bin sample galaxies and
construct $composite$ galaxies. To construct a $composite$ galaxy,
the member galaxies must have similar physical properties. We assume
that, if a set of galaxies have similar concentrations
(morphological type), physical sizes (angular momentum\footnote{At
given magnitude, the galaxies with larger angular momentum
(parameterized by larger spin parameter $\lambda$) tend to have
larger sizes \citep{mo98, syer99}.}) and luminosities (stellar
mass), their other physical properties are also similar, including
the radial color profiles.

We divide the 25,517 sample galaxies into subsamples in small
volumes of concentration $c$, physical half-light radius $R_{50}$
(in unit of kpc) and absolute Petrosian magnitude $M$. The bin
widths of $c$, $R_{50}$ and $M$ are 0.1, 1.0\,kpc and 0.5\,mag,
respectively.  The concentration is defined as $c\equiv
R_{90}/R_{50}$, where $R_{90}$ and $R_{50}$ are aperture radii
containing 90 and 50 percent of Petrosian flux (see EDR paper),
respectively. The physical size and absolute magnitude are
calculated under the cosmology contex with $\Omega_0=0.3$,
$\Omega_\Lambda=0.7$ and $H_0=70\rm{~km~s^{-1}Mpc^{-1}}$.

After dividing of galaxies into subsamples, we measure the
dispersion of the colors of all the member galaxies in each subset.
This dispersion is very small, with typical value $\sim0.1$~mag
(comparable to the bin width of the absolute magnitude), which
confirms our assumption that all the physical properties of member
galaxies are similar when $c$, $R_{50}$ and $M$ are constrained.

Since the member galaxies of each subset are assumed to have similar
color profiles but located at different redshifts, their color
profiles (measured from $profMean$ in a series of fixed angular
sizes) could be treated as the color measurement of `one' galaxy at
different physical radii. Therefore, the color profile of this
$composite$ galaxy could be constructed by combining color profiles
of all member galaxies in a subset (e.g., \citealt{azzollini08,
bakos08}).

Before make the composite color profiles, we have made a global
K-correction on the color profile of each member galaxy using the
routine {\sc kcorrect-v.4.1.4} \citep{blanton03a, blanton07}. In
principle, for each member galaxy, if color gradient exists, the
K-corrections at different radii will be different. This
second-order effect has been tested by using the $profMean$ data of
5 SDSS wavebands to make K-corrections for each photometric annulus
independently. We found the difference between the radius dependent
K-correction and the global one is less than $0.001$~mag.

\begin{figure}[t!!!]
 \centering
 \includegraphics[width=0.7\textwidth,angle=0]{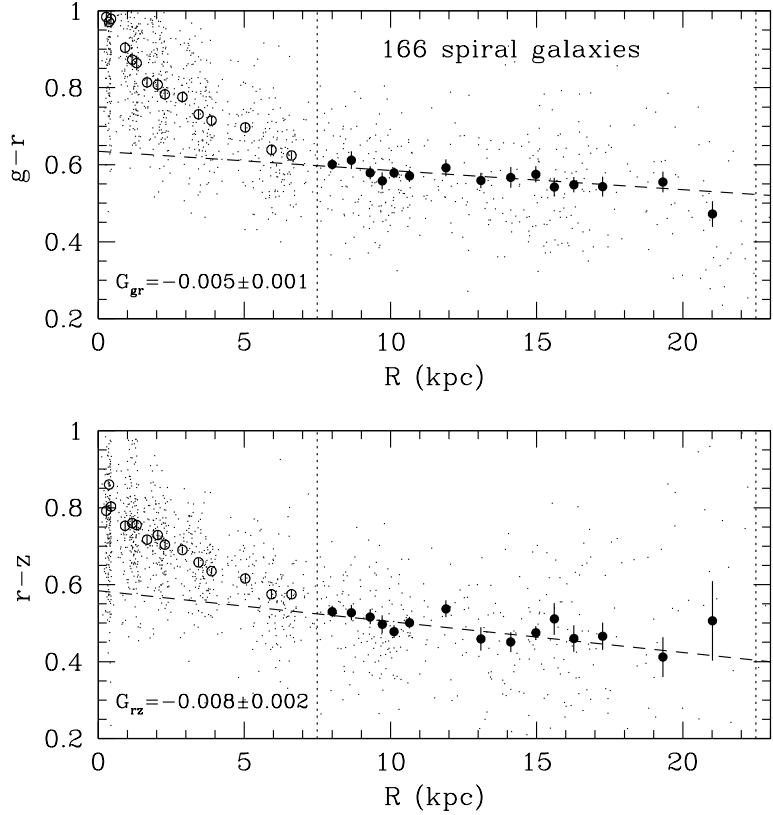}\\
\caption{The $g-r$ (upper panel) and $r-z$ (lower panel) color
profiles of an example $composite$ galaxy which is composed of 166
member galaxies with $1.9<c<2.0$, $7.0<R_{50}<8.0\,\rm{kpc}$ and
$-21.5<M<-21.0\,\rm{mag}$. Each small dot denotes the color at the
radius of photometric annulus for a member galaxy. The dashed line
is the best linear fitting of all the small dots between
$R_{50}\sim7.5\rm\,kpc$ and $3R_{50}$ (denoted by two vertical
dotted lines). The large circles show the average color at given
radius bins, with open and filled circles separated at $R_{50}$. The
error-bars show the scatters.}\label{fig:cp}
\end{figure}

Fig.~\ref{fig:cp} shows $g-r$ and $r-z$ color profiles of an example
$composite$ galaxy which comprises 166 member galaxies within the
volume of $1.9<c<2.0$, $7.0<R_{50}<8.0\,\rm{kpc}$ and
$-21.5<M<-21.0\,\rm{mag}$. In this figure, each small dot denotes
the color in a $profMean$ annulus of one member galaxy. These small
dots are then divided into 30 bins, with 15 bins inside $R_{50}$ and
the others outside. Each inside bin ($R<R_{50}$) contains the same
number of data points (average number of all small dots inside
$R_{50}$), so do the outside bins ($R>R_{50}$). We average the
colors of small dots in each bin and print them with large circles,
with open and filled ones representing inside and outside $R_{50}$
respectively. As we can see, by averaging a group of member
galaxies, the color profile of the $composite$ galaxy tends to be
smooth in the sense that the S/N of color profile has been greatly
improved.

To request high S/N of the stacked profile, we require the member
galaxies in each $c$-$R_{50}$-$M$ volume to be at least 30. With
this constraint, we get 193 $composite$ galaxies, containing
$\sim80$ percent of the initial 25,517 individual galaxies. Those
missed galaxies are located in the outskirts of the $c$-$R_{50}$-$M$
space with very low number density.

\subsection{The color gradient of disk component}\label{sec:data-cg-cg}

As shown in Fig.~\ref{fig:cp}, the color of this $composite$ galaxy
becomes bluer with increasing radius. The color profile is steeper
in the central part and becomes shallower outwards. The outer part
of color profile ($R>R_{50}$) shows a nearly linear shape.

Following the study of \citet{taylor05}, we separate the color
profile of each $composite$ galaxy into two parts at $R_{50}$ and
consider the region $R>R_{50}$ as the disk component. We measure the
color gradient of disk component within the region from $R_{50}$ to
$3R_{50}$. The outer boundary is based on the considerations that a
star formation threshold occurs outside the radius of about
$5$-times of the scale-length $R_{d}$ ($\sim3R_{50}$)
\citep{vanderkruit81, pohlen00} and the large uncertainties of the
photometric measurements exist in the very outer
regions\footnote{The typical errors of surface brightness at
$3R_{50}$ are $0.105$ for $g$-band, $0.095$ for $r$-band and $0.273$
for $z$-band, respectively. }.

Since the color profile of disk component ($R_{50}\sim3R_{50}$) can
be well parameterized by a straight line, we use the linear least
squares fitting to model the gradients of galactic disks. In
specific, we make a linear fitting of all the data points in the
range $R_{50}\sim3R_{50}$ and estimate the fitting error with 20
bootstrap samples. The best linear fitting for the example galaxy is
shown as the dashed line in Fig.~\ref{fig:cp}.

The disk color gradient is then defined as the slope of the fitting
line,
\begin{equation}
 G_{\rm{kpc}} = \frac{\Delta CI}{\Delta R (\rm kpc)}\label{eq:Gkpc}\\
\end{equation}
where $CI$ is the color index. For our $composite$ galaxy, this
$G_{\rm kpc}$ can be easily converted to the scaled color gradient,
e.g. $G_{50}$ (the color variation per $R_{50}$), by
$G_{50}=G_{\rm{kpc}}\times R_{50}$, where $R_{50}$ is the average
size of member galaxies in each subset.

In this paper, we will only present the results of colors $g-r$ and
$r-z$. We exclude the colors related to $u$ band because of its low
image sensitivity and high sensitivities to recent star formation
and dust attenuation. We do not use $i$-band because of the `Red
halo' effect of the Point Spread Function (PSF), which was
introduced by \citet{michard02} and was first found by \citet{wu05}
in SDSS $i$-band. Except for the $i$-band, the PSF profiles in $g$,
$r$ and $z$ bands match well with each other \citep{wu05}, so the
effects of the PSF mismatch on the $g-r$ and $r-z$ color profiles
are negligible.

\section{RESULTS}\label{sec:result}

\subsection{Color gradient}\label{sec:result-cg}

\begin{figure}[!htbp]
 \centering
 \includegraphics[width=0.8\textwidth]{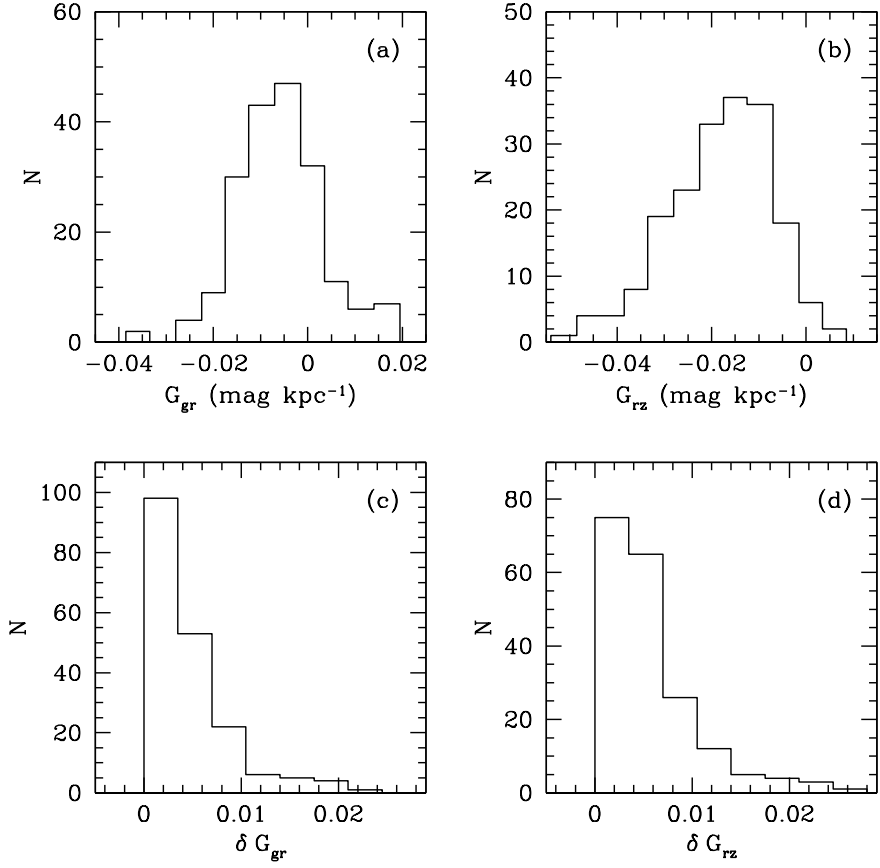}\\
\caption{The distributions of color gradients $G_{gr}$ [panel (a)]
and $G_{rz}$ [panel (b)] of $composite$ galaxies. The bottom two
panels show the distributions of errors of
$G_{gr}$[$\;\delta\,G_{gr}$, panel (c)] and
$G_{rz}$[$\;\delta\,G_{rz}$, panel (d)] respectively.}
\label{fig:hist-cg}
\end{figure}

We measure the color gradients of disks ($R_{50}\sim3.0R_{50}$) of
all 193 $composite$ galaxies and show their distributions in
Fig.~\ref{fig:hist-cg}. Panels (a) and (b) show the histograms of
$g-r$ gradient $G_{gr}$ and $r-z$ gradient $G_{rz}$, whereas panels
(c) and (d) show the histograms of their errors, respectively.

Consistent with many early studies (e.g., \citealt{deJong96b,
bell00, moth02, taylor05, li05, bakos08}), most of our $composite$
galaxies show negative color gradients. We notice that, for $g-r$
color, about 25 percent ones have positive gradients. As we will
show below, these $composite$ galaxies with positive $G_{gr}$ are
mainly those with small $R_{50}$ and large uncertainties.

As can be seen in Fig.~\ref{fig:hist-cg}, both $G_{gr}$ and $G_{rz}$
show roughly Gaussian distributions with means and scatters
$\bar{G}_{gr}=-0.006\rm\,mag\,kpc^{-1}$,
$\bar{G}_{rz}=-0.018\rm\,mag\,kpc^{-1}$ and $\sigma_{gr}=0.009$,
$\sigma_{rz}=0.011$. The typical errors of fitted gradients are
about $0.003$ for $G_{gr}$ and $0.005$ for $G_{rz}$, which are much
smaller than the scatters $\sigma_{gr}$ and $\sigma_{rz}$. If we
de-convolve these typical errors from the distributions of $G_{gr}$
and $G_{rz}$, we get intrinsic scatters $\sigma_{gr,\rm\,i}=0.008$
and $\sigma_{rz,\rm\,i}=0.010$. These non-negligible intrinsic
scatters among different $composite$ galaxies imply that the color
gradient is intrinsically correlated with the physical properties of
galaxies. We discuss  possible correlations in detail below.

\subsection{Correlation with galaxy properties}\label{sec:result-relate}

In this section, we study the correlations between the color
gradients and structural parameters of galaxies and try to find out
which one is the main factor regulating the color gradients.
Technically, we use the Pearson's correlation coefficient to
quantify each possible correlation\footnote{During the calculation
of this coefficient, the error of the color gradient of each
$composite$ galaxy has been used as a weight.}. In the case of our
sample containing 193 data points, the confidence level of the
correlation will be over 99 percent if the absolute value of a
correlation coefficient is larger than 0.18.

\begin{figure}[h!!!]
 \centering
 \includegraphics[height=\textwidth,angle=0]{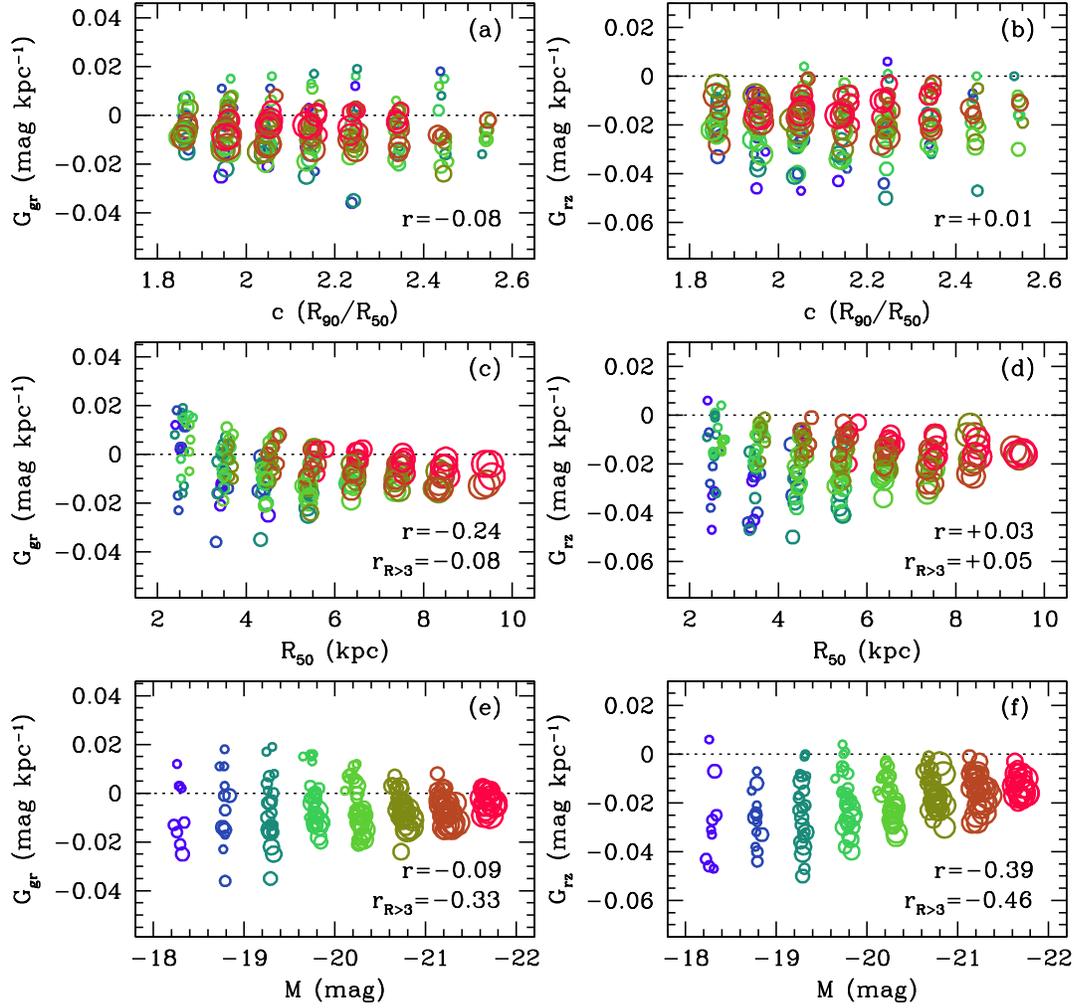}\\
\caption{The color gradients ($G_{gr}$ and $G_{rz}$) as functions of
concentration parameter $c$, physical half-light radius $R_{50}$ and
absolute magnitude $M$. The correlation coefficients $r$ are labeled
in the lower-right corner of each panel. Each point represents the
color gradient of one $composite$ galaxy. Bigger circles denote
galaxies with larger sizes and redder colors represent higher
luminosities. } \label{fig:cg-3}
\end{figure}

As described in section~\ref{sec:data-cg-bin}, our $composite$
galaxies are constructed in $c$-$R_{50}$-$M$ space, so we explore
the possible correlations between color gradients and these three
bin parameters first. Fig.~\ref{fig:cg-3} shows the measured color
gradients ($G_{gr}$ and $G_{rz}$) against the $r$-band concentration
$c$, physical size $R_{50}$ and absolute magnitude $M$ in the top,
middle and bottom panels, respectively.  Each circle in the figure
represents one $composite$ galaxy, with sizes and colors of the
circles indicating  $R_{50}$ and $M$ of the galaxies. Larger circles
represent larger galaxies and redder colors indicate more luminous
galaxies. The errors of the color gradients are not shown for
clearness (but shown in Fig.~\ref{fig:cg-mu}). The Pearson's
correlation coefficients between color gradients and these three
parameters ($c$, $R_{50}$ and $M$) are labeled in the corner of each
panel and listed in Table~\ref{tab:correlation}.

\begin{table}[h!!!]
 \small
 \centering
 \caption{The Pearson's correlation coefficients between color
gradients ($G_{gr}$ and $G_{rz}$) and four structural parameters
($r$-band): concentration parameter $c$, physical half-light radius
$R_{50}$, absolute magnitude $M$ and disk surface brightness
$\mu_{d}$.}\label{tab:correlation}
 \begin{tabular}{lcccc}
 \hline\noalign{\smallskip}
  & $c$
  & $R_{50}$
  & $M$
  & $\mu_{\rm d}$ \\
 \hline\noalign{\smallskip}
 $G_{gr}$ & $-0.08$ & $-0.24$ $(-0.08^{*})$ & $-0.09$ $(-0.33^{*})$ & $-0.71$ \\
 $G_{rz}$ & $0.01$  & $0.03$  $( 0.05^{*})$ & $-0.39$ $(-0.46^{*})$ & $-0.75$\\
 \noalign{\smallskip}\hline
\end{tabular}
\tablecomments{0.8\textwidth}{* -- The correlation coefficients of
$composite$ galaxies with $R_{50}>3$~kpc.}
\end{table}

\subsubsection{With concentration parameter $c$}\label{sec:result-relate-c}
As we can see from the top two panels of Fig. 4, the color gradients
are almost independent of the concentration parameter. The
correlation coefficients $r_{G_{gr},\,c}=-0.08$ and
$r_{G_{rz},\,c}=0.01$ confirm that the color gradients are not
correlated with $c$.

It is well known that the concentration parameter is related to the
morphological type, with earlier type galaxies having larger values
of $c$ \citep{shimasaku01, strauss02, nakamura03}. For spiral
galaxies, a larger value of $c$ also indicates an earlier type
spiral, i.e. larger contribution from bulge component. Therefore,
our results imply that the color gradient of the disk component is
not related to the bulge. Moreover, this independence of the bulge
component also confirms that our separation of disk component at
$R_{50}$ is adequate to avoid the bulge contamination.

\subsubsection{With physical size $R_{50}$}\label{sec:result-relate-r}
For physical size $R_{50}$, as can be noticed from the correlation
coefficients, the gradient $G_{gr}$ shows a weak correlation
($r_{G_{gr},\,R_{50}}=-0.24$), while $G_{rz}$ is independent of
$R_{50}$ ($r_{G_{rz},\,R_{50}}=0.03$). If we go to the plot of
$G_{gr}~vs~R_{50}$ [panel $(c)$ of Fig.~\ref{fig:cg-3}], we can see
that the weak correlation between $G_{gr}$ and $R_{50}$ is mainly
caused by the galaxies in the smallest size bin, where more than
half of the $composite$ galaxies show positive gradients. Actually,
if we exclude these galaxies from the correlation analysis, we get a
coefficient $r=-0.008$, a similar independent behavior as that of
the $r-z$ gradient. Meanwhile, the correlation coefficient
$r_{G_{rz},\,R_{50}}$ changes from $0.03$ to $0.05$. We also mention
that the measurement of the color gradients of these smallest
galaxies is quite uncertain (see the error-bars of these smallest
galaxies in Fig.~\ref{fig:cg-mu}) because of the limited number of
member galaxies and $profMean$ data points. However, the fact that
the smallest galaxies show positive $G_{gr}$ but negative $G_{rz}$
may relate to the migrations of star-burst regions in blue compact
galaxies (e.g., \citealt{Schulte99, depaz05}) and worths further
studying.

Another interesting point we can see from the figure is that, the
dispersion of the color gradients (for both $G_{gr}$ and $G_{rz}$)
tends to be smaller for bigger galaxies. For the biggest galaxy
($R_{50}\sim9\rm\,kpc$), the gradients roughly keep as constants
$G_{gr,\,9\rm kpc}\sim-0.008$ and $G_{rz,\,9\rm kpc}\sim-0.016$
(where the dispersion is comparable to measurement error). Similar
result has also been found by \citet{munoz07}, who investigated 161
nearby face-on spiral galaxies and found smaller $FUV-K$ gradient
dispersion for bigger galaxies.

\subsubsection{With absolute magnitude $M$}\label{sec:result-relate-m}
The correlation coefficients between the color gradients and
absolute magnitude are $r_{G_{gr},\,M}=-0.09$ and
$r_{G_{rz},\,M}=-0.39$. Again, when we exclude the composite
galaxies in smallest size bin as above, the coefficient
$r_{G_{gr},\,M}$ increases to $-0.33$ and $r_{G_{rz},\,M}$ becomes
$-0.46$. The weak anti-correlation between the color gradient and
absolute magnitude indicates that brighter galaxies have shallower
gradients. This result is apparently conflict with the result of
\citet{kim90}, where they found more luminous galaxies tend to have
steeper color gradients. However, this apparent contradictory is
caused by different definitions of the color gradients.
\citet{kim90} defined the color gradient as color variation per
scaled radius $D_{25}$\footnote{\citet{kim90} defined the $D_{25}$
as isophotal diameter for $\mu=25$\,mag\,arcsec$^{-2}$ isophote in
$B$-band.}, while we adopt the definition of color variation per
kpc. As we have mentioned, the color gradient $G_{\rm kpc}$ can be
converted to the scaled  color gradient by multiplying a factor of
physical radius. Since more luminous galaxies have larger sizes
\citep{shen03}, if we convert $G_{\rm kpc}$ to $G_{50}$, we also get
a positive correlation between $G_{50}$ and absolute magnitude with
coefficients $0.17$ and $0.18$ for $g-r$ and $r-z$ gradient
respectively, i.e. the more luminous galaxies have steeper scaled
color gradients.

Till now, we have explored the correlations between the color
gradients and three bin parameters separately and found a tendency
of lower scatter of color gradients for larger galaxies and a weak
correlation between the color gradient and absolute magnitude. If we
combine these bin parameters together, we may find more interesting
results. In the middle two panels of Fig.~\ref{fig:cg-3}, if we
focus on symbol colors, we will find a systematical change that less
luminous galaxies (with bluer symbol colors) tend to have steeper
color gradients in a given radius bin. Similar systematical change
can also be found in bottom panels of Fig.~\ref{fig:cg-3}, where
larger galaxies (with bigger symbol sizes) have steeper color
gradients at given absolute magnitude. These two phenomena indicate
a strong correlation between the color gradients and surface
brightness. We discuss this correlation in detail below.

\subsubsection{With disk surface brightness}\label{sec:result-relate-u}

In this section, we turn to investigate the correlation between the
color gradients and disk surface brightness $\mu_{d}$. Here,
$\mu_{d}$ is defined as average surface brightness from $R_{50}$ to
$R_{90}$ and has been corrected by K-correction and cosmological
dimming effect. This definition of the disk surface brightness
avoids  possible contaminations from bulge component and is
consistent with our measurement of the disk color gradient. We
calculate $\mu_{d}$ for each member galaxy and then average them to
get the disk surface brightness of the $composite$ galaxy.

\begin{figure}[h!!!]
 \centering
 \includegraphics[width=0.8\textwidth]{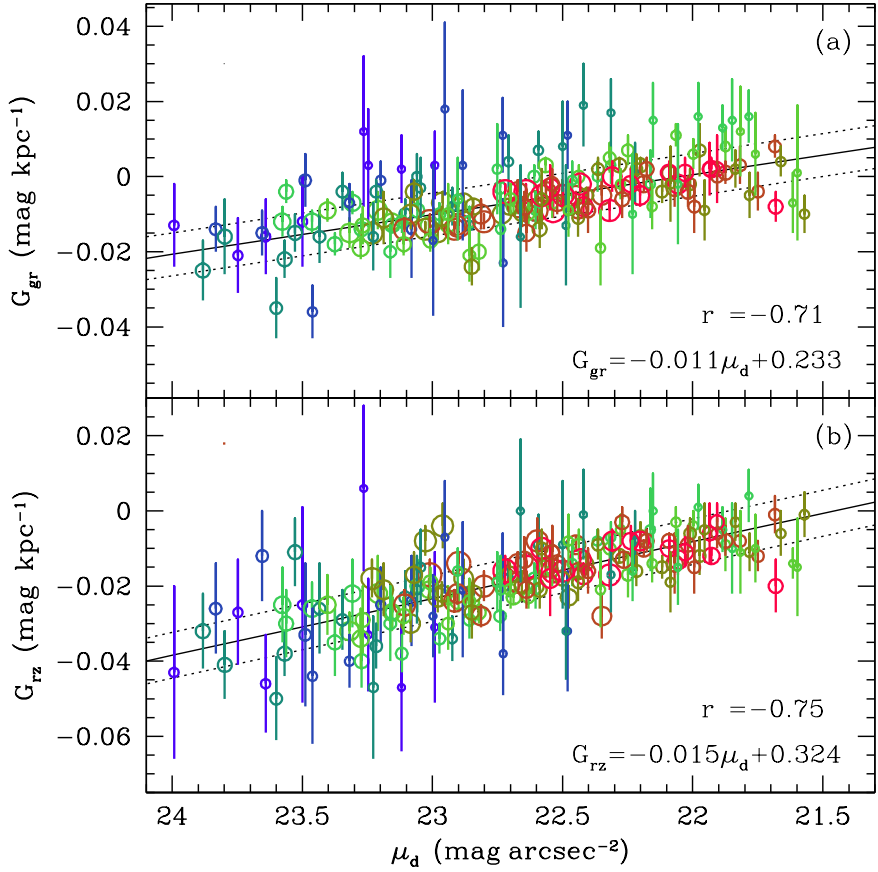}\\
\caption{The relationship between color gradients ($G_{gr}$ and
$G_{rz}$) and disk surface brightness $\mu_{d}$. Each point
represents the color gradient of one $composite$ galaxy and the
error-bar shows the measurement error estimated from bootstrap
samples. The solid line in each panel shows the best linear fitting
of all the data points and the dotted lines present the region of
$\pm1\sigma$ $rms$ scatter.} \label{fig:cg-mu}
\end{figure}

We plot the color gradients against disk surface brightness in
Fig.~\ref{fig:cg-mu}. Both $G_{gr}$ and $G_{rz}$ show a remarkable
anti-correlation with $\mu_{d}$, with correlation coefficients
$r_{G_{gr},\,\mu_{d}}=-0.71$ and $r_{G_{rz},\,\mu_{d}}=-0.75$. We
use the least squares method to fit linear relations between the
color gradients and $\mu_{d}$ weighted with the errors of the color
gradients. The best fitting results are
\begin{equation} \label{fit1}
 \begin{aligned}
   G_{gr} &= -0.011\mu_{d} + 0.233 ,\\
   G_{rz} &= -0.015\mu_{d} + 0.324 ,
 \end{aligned}
\end{equation}
which are shown as the solid lines in Fig.~\ref{fig:cg-mu}. The
$rms$($\sigma$) of these fitted relations are 0.005 for $G_{gr}$ and
0.006 for $G_{rz}$. The range of the $\pm1\sigma$ region is shown as
two dotted lines in each panel. As we can see, most of the galaxies
outside the $\pm1\sigma$ region are small galaxies (small symbol
size) with large measurement errors. Moreover, we have tested that
the residuals of these best fitting relations are not correlated
with either the luminosities or the sizes of galaxies (see the
colors and sizes of symbols). That means, the weak correlation
between the color gradients and absolute magnitude we have found in
Section~\ref{sec:result-relate-m} is actually caused by this
$G-\mu_d$ relation and known $L-\mu_d$ relation (e.g.,
\citealt{shen03}). This also indicate that the tight correlation
between color gradients and disk surface brightness is an intrinsic
relation of spiral galaxies, which gives a strong observational
constraint on the models of formation and evolution of galactic
disk.

\section{SUMMARY AND DISCUSSION}\label{sec:dis}

In this study, a sample of 25,517 face-on spiral galaxies drown from
SDSS-DR4 is used to investigate the color gradients of galactic
disks. We construct $composite$ galaxies to improve the S/N of color
profiles and measure their color gradients, and then investigate the
correlations between the color gradients and structural parameters,
including concentration, physical size, luminosity and disk surface
brightness. The main results are listed as follows.

\begin{enumerate}

\item The color profile of disk component, from $R_{50}$ to $3R_{50}$, can
be well represented by a straight line. Almost all of $composite$
galaxies show negative $r-z$ gradient, while about 75 percent of
them show negative $g-r$ gradients. The mean color gradients are
$\bar{G}_{gr}=-0.006$\,mag\,kpc$^{-1}$ and
$\bar{G}_{rz}=-0.018$\,mag\,kpc$^{-1}$, respectively.

\item The color gradients of galactic disks are independent of
morphological types (concentration $c$).

\item There is also no significant correlation between disk color
gradients and the size of galaxies ($R_{50}$), but bigger galaxies
tend to have smaller scatters and uncertainties of the color
gradients.

\item The color gradients are dependent on the absolute magnitude
$M$, with fainter galactic disks having steeper color gradients. But
this correlation may caused by $G-\mu_d$ relation.

\item The most remarkable correlation we find is between the
color gradients and disk surface brightness $\mu_{d}$, which can be
fitted by straight lines: $G_{gr}=-0.011\mu_{d}+0.233$ and
$G_{rz}=-0.015\mu_{d}+0.324$.
\end{enumerate}

There are two potential causes of color gradients in galactic disks.
One is the systematical change of dust attenuation along the disk
and the other is the gradient of stellar population.

For dust, a reasonable and simplified assumption is that the dust
distribution along the disk is proportional to the stellar mass
density profile (e.g., \citealt{phillipps88, regan06}). Thus, the
inner disk contains more dust than outer region, and naturally
generate a negative color gradient (redder inward) as we observed.
However, in this scenario, high surface brightness (surface mass
density) galaxies have more dust and will generate steeper color
gradients than low surface brightness galaxies, which is opposite to
our observational results. Therefore, dust attenuation could not be
the only role in the origin of color gradients of galactic disks.
Many previous works also have suggested that the dust only plays a
minor role in regulating the color gradients of spiral galaxies
(e.g., \citealt{deJong96c, bell00, macArthur04, taylor05}).

Regarding the stellar population, it is related to chemical
evolution history of the spiral galaxies and has only been studied
in details for local group galaxies (e.g., MW: \citealt{chang99,
chen03, fu09}; M31: Yin et al. in preparation; M33:
\citealt{magrini07, barker08}). It is broadly accepted that the star
formation rate (SFR) is proportion to the surface mass density
(e.g., \citealt{bell00, kauffmann03}), and may also inverse
proportion to radius (e.g., \citealt{kennicutt98, boissier99}),
which means SFE (star formation efficiency, defined as
$\textrm{SFR}/\Sigma_{gas}$, where $\Sigma_{gas}$ is gas surface
density) is higher in the inner region than outer part. Thus, in the
inside-out star-forming scenario, the inner regions of galactic
disks evolved faster than outer regions due to higher SFE, leading
to older and more metal-rich stellar populations and generating
negative color gradients as observed (\citealt{prantzos00}, Yin et
al. in preparation). Moreover, the Kennicutt star formation low is
nonlinearly proportional to the surface brightness as
$\textrm{SFR}\propto\mu^{1.4}$ \citep{kennicutt98}. So, the higher
surface brightness galaxies have higher SFE, i.e. the chemical
evolution will be more adequate. \citet{boissier00} have modeled the
color profiles of galactic disks using a nonlinear star formation
law and found that more massive and smaller galaxies (i.e., higher
surface density ) favor a more rapid early star formation than the
lower surface density ones. Their gas fractions are lower and
chemical evolutions are more adequate nowadays even in the outer
region, which means the abundance and stellar population vary with
radius slowly. Therefore the color differences between the inner and
outer regions of high surface brightness galaxies will be smaller
than that of low surface brightness ones. This simple argument
concludes a consistent trend with our observational results that the
higher surface brightness galaxies have shallower color gradients.
However, besides this static chemical evolution, the growth and
evolution of galactic disks are related to a lot of complicated
processes, e.g. gas cooling, outflow, radial mixing, minor merging
(e.g., \citealt{kaiser03, block06}). To fully understand and
quantify the origin and evolution of color gradients observed in
spiral galaxies, more detailed models are required.

\normalem
\begin{acknowledgements}
The authors thank Bo Zhang, Junliang Zhao, Huanan Li and Quanbao
Xiao for helpful discussions and suggestions. This work is partly
supported by National Science Foundation of China with No. 10573028
and 10803016, the Key Project with No. 10833005 and 10878003, the
Group Innovation Project with No. 10821302, 973 program with No.
2007CB815402 \& 403, and the Knowledge Innovation Program of the
Chinese Academy of Science.

Sponsored by Shanghai Rising-Star Program(Shen)

Funding for the creation and distribution of the SDSS Archive has
been provided by the Alfred P. Sloan Foundation, the Participating
Institutions, the National Aeronautics and Space Administration, the
National Science Foundation, the U.S. Department of Energy, the
Japanese Monbukagakusho, and the Max Planck Society. The SDSS Web
site is http://www.sdss.org/.

The SDSS is managed by the Astrophysical Research Consortium (ARC)
for the Participating Institutions. The Participating Institutions
are The University of Chicago, Fermilab, the Institute for Advanced
Study, the Japan Participation Group, The Johns Hopkins University,
the Korean Scientist Group, Los Alamos National Laboratory, the
Max-Planck- Institute for Astronomy (MPIA), the Max-Planck-Institute
for Astrophysics (MPA), New Mexico State University, University of
Pittsburgh, University of Portsmouth, Princeton University, the
United States Naval Observatory, and the University of Washington.
\end{acknowledgements}



\label{lastpage}

\end{document}